\documentclass [12pt]{article}
\begin{document}

\begin{center}
\Large{\bf Infrared Consistency of NSVZ and DRED Supersymmetric Gluodynamics} 
\end{center}

\bigskip

\begin{center}
V. Elias\\
Department of Applied Mathematics\\
University of Western Ontario\\
London, Ontario N6A 5B7\\
Canada
\end{center}

\begin{abstract}
Pad\'e approximant methods are applied to the known terms of the DRED
$\beta$-function for $N=1$ supersymmetric SU(3) Yang-Mills theory.  Each of the $[N|M]$
approximants with $N+M \leq 4$, $M \neq 0$ constructed from this series
exhibits a positive pole which precedes any zeros of the approximant,
consistent with the same infrared-attractor pole behaviour known to
characterize the exact NSVZ $\beta$-function.  A similar 
Pad\'e-approximant analysis of truncations of the NSVZ series is
shown consistently to reproduce the geometric-series pole of the exact
NSVZ $\beta$-function.
\end{abstract}

\section*{1.  Introduction: Supersymmetric Gluodynamics and the Adler-Bardeen Theorem}

Nearly two decades ago, Novikov, Shifman, Vainshtein, and Zakharov (NSVZ) derived the exact 
$\beta$-function for supersymmetric gauge theories with no matter fields [1].  
However for the $N=1$ case, it was recognized early-on that scheme dependence characterizes higher-than-three 
loop terms of this  $\beta$-function; a prior three-loop calculation using now-standard 
dimensional reduction (DRED) methods \cite{ACV} was seen to yield a $\beta$-function that differs from the 
NSVZ result (as anticipated in \cite{DRT}) at three-loop-order.  The present work seeks to address whether this discrepancy 
is indicative of differences more profound than those which can be accommodated by 
order-by-order redefinition of the coupling constant.

	For simplicity, we will focus our attention on $N=1$ supersymmetric SU(3) 
Yang-Mills theory (supersymmetric gluodynamics).  Arguments advanced by Jones \cite{DRT}, 
which are recapitulated below, can be used to extract the exact
NSVZ $\beta$-function for this theory.  Within supersymmetric
gluodynamics, the supermultiplet structure of the anomalies entails the
following relation:

\renewcommand{\theequation}{1.\arabic{equation}}
\setcounter{equation}{0}  

\begin{equation}
\partial_\mu J_5^\mu (x) = \partial_\mu \left[ \frac{1}{2} \bar{\lambda}
(x) \gamma^\mu \gamma_5 \lambda (x) \right] = \frac{\beta (g)}{g} \cdot
\frac{4}{3} G(x)
\end{equation}
$[\beta(g) = \mu dg / d\mu]$ where $\lambda(x)$ are gluino fields, and where
\begin{equation}
G = - \frac{1}{8} F \tilde{F} + \frac{1}{4} \partial_\mu \left[
\bar{\lambda}(x) \gamma^\mu \gamma_5 \lambda(x) \right]
\end{equation}
is a component of the Wess-Zumino chiral supermultiplet
(A,B,$\psi$,F,G).  Eq. (1.2) may be rearranged to read
\begin{equation}
F \tilde{F} = -8 \left[ G - \frac{1}{2} \partial_\mu J_5^\mu \right].
\end{equation}
However, the Adler-Bardeen theorem requires for $N_c = 3$ that
\cite{DJL}
\begin{equation}
\partial_\mu J_5^\mu = \frac{1}{2} \frac{3 g^2}{16 \pi^2} F \tilde{F}.
\end{equation}
By substituting (1.3) into the right-hand side of the (1.4), one finds
that
\begin{equation}
\partial_\mu J_5^\mu = - \frac{3 g^2}{4 \pi^2} \left[ G - \frac{1}{2}
\partial_\mu J_5^\mu \right]
\end{equation}
in which case
\begin{equation}
\partial_\mu J_5^\mu (x) = \left[ (-3 g^2 / 4 \pi^2) / (1 - 3 g^2 / 8
\pi^2) \right] G(x).
\end{equation}
Comparing the right-hand sides of (1.1) and (1.6), one finds that
\begin{equation}
\beta (g) / g = -\frac{9g^2 / 16 \pi^2}{1 - 3g^2 / 8 \pi^2}.
\end{equation}
This $\beta$-function manifestly acquires a pole at $\alpha = 2 \pi /
3$, and indeed corresponds to the NSVZ $\beta$-function
whose infrared properties are discussed by Kogan and Shifman
in ref. \cite{KS}.  In other words, the exact NSVZ $\beta$-function for supersymmetric
gluodynamics can be extracted by imposing the Adler-Bardeen theorem on
the supermultiplet structure of the anomalies \cite{DRT}.

Alternatively, the analysis of ref. \cite{DRT} can be turned around to show that the NSVZ $\beta$-function for supersymmetric
gluodynamics, as derived in \cite{NSV} via instanton calculus methods, is consistent to all orders with the
Adler-Bardeen theorem.  By contrast, the DRED $\beta$-function, which
differs from the NSVZ $\beta$-function subsequent to identical leading-
and next-to-leading-order contributions, is necessarily
{\it inconsistent} with the Adler-Bardeen theorem after next-to-leading order.  

The question of
whether or not the DRED $\beta$-function is therefore ``wrong'' can be sidestepped
by a perturbative redefinition of the coupling \cite{JJN,IJD}.  Denoting 
by $x$ and $y$ the corresponding NSVZ- and DRED-coupling parameters ($x \equiv g^2 / 16
\pi^2$), the respective perturbative $\beta$-functions taken to three
non-leading orders \cite{IJD}
are consistent with the coupling constant redefinition \footnote{In the notation 
of ref. \cite{IJD}, $(g^{DRED})^2 = (g^{NSVZ})^2 + \delta^{(1)} + \delta^{(2)} + \delta^{(3)}$.
If $n_f = 0$ and $N_c = 3$, the ref. \cite{IJD} group-theoretical
factors for supersymmetric gluodynamics are $C(R) = T(R) = 0, \; \; C(G)
= 3$, and $Q = -9$, in which case $\delta^{(1)} = 0, \; \; \delta^{(2)}
= 27 x^2 (g^{NSVZ})^2$, and $\delta^{(3)} = 351 x^3 (g^{NSVZ})^2$.}
\begin{equation}
y=x \left[1+27x^2 + 351x^3 + ... \right] .
\end{equation}
However, the deeper question of validity ultimately rests upon the
consistency of the non-perturbative content of the DRED and NSVZ schemes.
In ref. \cite{JJN}, for example, the suggestion is made that a relation may be found between
possible infrared fixed points of $\beta$-functions in both schemes when matter fields are present.  Our
focus here, though, will be on $\beta$-function properties in the {\it absence} of matter fields,
since these properties are fully accessible for the NSVZ case.

In the NSVZ scheme, the $\beta$-function (1.7) leads to a double-valued couplant.
The pole at $\alpha = 2 \pi/3 \; (x=1/6)$, is an infrared-attractor point
common to both phases of the couplant \cite{KS}, {\it i.e.}, an
asymptotically-free phase ($x < 1/6$ and $x \rightarrow 0$ as $\mu
\rightarrow \infty$) and a non-asymptotically free strong phase (x {\it increases} monotonically 
with $\mu$ and is greater than 1/6). Thus, the infrared
content of supersymmetric gluodynamics is governed by a {\it pole}, as opposed
to an infrared fixed point corresponding to a positive $\beta$-function zero.

It would clearly be of value to determine whether the same infrared
behaviour characterizes the DRED $\beta$-function for supersymmetric
gluodynamics. However, the DRED $\beta$-function series is presently
known only to four-loop order.  Consequently, we choose to utilize 
Pad\'e-approximant methods that are already known to be of value in extracting
poles of functions underlying truncated power series \cite{BG}.  In
Section 2, we test this approach by examining whether Pad\'e-approximants constructed from
truncations of the geometric NSVZ $\beta$-function series succeed in
reproducing the infrared-attractor pole characterizing the full 
$\beta$-function's infrared content.  We find this to be the case for all $M
\neq 0$ $[N|M]$ approximants that can be constructed
from the first five terms of the NSVZ $\beta$-function series (the $M=0$ case
is the truncated series itself).  We also
find that asymptotic Pad\'e-approximant-prediction (APAP) formulae for
subsequent terms of a truncated geometric series ({\it e.g.}, truncations of the
NSVZ $\beta$-function series) are fully consistent with that series
remaining geometric to next order.

In Section 3, the same Pad\'e-approximant procedures are applied to
known terms of the DRED $\beta$-function series. We reconfirm the
observation \cite{IJD} that APAP methods are successful (within 10\%
accuracy) in predicting the four-loop term of this series from its first
three terms.  We find that all $[N|M]$ approximants constructed from the
four (or fewer) known terms of the DRED $\beta$-function series $(N+M \leq 3)$
are characterized by a positive pole which precedes any positive zeros.  
Moreover, if these four known terms are augmented by the APAP prediction for the five-loop
term of the series, this same result is found to be applicable for the
$N+M=4$ case of $[N|M]$ approximants to the five-loop series.

Such a pole common to {\it all} $N+M \leq 4$ approximants to the DRED $\beta$-
function is suggestive of an infrared-attractor pole for the true DRED
$\beta$-function.  However, the magnitude of the pole, as indicated by
the full set of $N + M \leq 4$ approximants, is consistently seen to be less than the
$x = 1/6$ value characterizing the NSVZ $\beta$-function.  This
contradicts expectations [{\it e.g.}, eq. (1.8)] that couplant values in the
DRED scheme should be {\it larger} than their corresponding values in
the NSVZ scheme.  In Section 4, this issue is examined in
detail.  In particular, the formula (1.8) is demonstrated to be a
special case of a more general formula which {\it does} permit DRED
couplant values  that are smaller than 
their corresponding NSVZ values for the asymptotically-free phase.
We thus conclude that the smaller infrared-attractor pole that appears
to characterize the DRED $\beta$-function is not necessarily
incompatible with the pole characterizing the NSVZ $\beta$-function for
supersymmetric gluodynamics.

\section*{2.  Pad\'e-Approximants and NSVZ Supersymmetric Gluodynamics}

\renewcommand{\theequation}{2.\arabic{equation}}
\setcounter{equation}{0}     

	We would like to probe the infrared structure of the DRED $\beta$-function for 
supersymmetric gluodynamics, and compare its infrared properties to those of the NSVZ 
$\beta$-function.  The NSVZ  $\beta$-function for supersymmetric gluodynamics 
($N=1$ supersymmetric $SU(3)$ Yang-Mills theory) is
\begin{equation}
\mu^2 \frac{dx}{d\mu^2} \equiv \beta^{NSVZ} (x) = -9x^2 / (1-6x),
\end{equation}
where
\begin{equation}
x \equiv \alpha^{NSVZ} (\mu) / 4 \pi = (g^{NSVZ} (\mu))^2 / 16 \pi^2.
\end{equation}
This $\beta$-function's infrared behaviour is governed by a pole at $x = 1/6$.  Values of the 
couplant between zero,
the ultraviolet fixed point, and $1/6$, the infrared-attractor pole, correspond to 
the asymptotically-free phase of supersymmetric gluodynamics, as discussed in \cite{KS}. The 
perturbative series for the NSVZ $\beta$-function is clearly geometric to all orders:
\begin{equation}
\beta^{NSVZ} (x) = -9x^2 \left[ 1 + 6x + 36x^2 + 216x^3 + ... \right].
\end{equation}

By contrast the corresponding DRED $\beta$-function is known only from the first four terms of its 
perturbative series \cite{ACV,IJD}:
\begin{equation}
\mu^2 \frac{dy}{d\mu^2} \equiv \beta^{DRED} (y) = -9y^2 \left[
1+6y+63y^2+918y^3 + ... \right],
\end{equation}
\begin{equation}
y \equiv \alpha^{DRED} (\mu) / 4\pi = (g^{DRED} (\mu))^2 / 16 \pi^2.
\end{equation}
In the absence of any further information about the DRED $\beta$-function, we necessarily have to 
make use of Pad\'e-approximant methods to explore whether the same infrared behaviour characterises the 
asymptotically-free phases of DRED and NSVZ versions of supersymmetric gluodynamics.

	The general problem of utilizing Pad\'e approximants to ascertain infrared properties 
of $\beta$-functions from their perturbative series
\begin{equation}
\beta(z) = \beta_0 z^k \left[ 1+R_1 z + R_2 z^2 + R_3 z^3 + R_4 z^4 +
... \right]
\end{equation}
is addressed in reference \cite{CEM}, and is an example of the use of  Pad\'e-approximant poles 
to probe the singularity structure of functions underlying the (known) terms of truncated series 
\cite{BG}. One self-evident criterion 
for such use of Pad\'e approximants is the requirement that Pad\'e-approximant predictions of 
$R_{N+1}$ (based on known values for $\{R_1, R_2, ..., R_N\}$) become progressively more accurate as $N$ 
increases. If such is the case, and if a leading zero (infrared fixed point) or pole 
(infrared-attractor point) {\it consistently} emerges from all possible approximants whose power
series reproduce the known coefficients $R_k$, then that leading structure (be it a zero or a pole) is 
unlikely to be a Pad\'e artifact ({\it e.g.}, a defect pole) such as might arise from only a single 
Pad\'e approximant to the truncated series \cite{BG}.

	Pad\'e approximant predictions of $R_{N+1}$ are based on $[N-M | M]$ Pad\'e approximants whose 
Maclaurin expansions reproduce the ``known'' terms $1 + R_1 z + R_2 z^2 + ... + R_N z^N$ of the series.  
For example, even two known terms $1 + R_1 z$ are sufficient to determine the $[0 | 1]$ Pad\'e 
approximant $1/(1 - R_1 z) = 1 + R_1 z + R_1^2 z^2 + ... $, which can then be said to predict the 
``unknown'' next-order coefficient $R_2$: $R_2^{[0 | 1]} = R_1^2$. If this prediction differs in sign or 
greatly in magnitude from the true calculated value of $R_2$, then there is little reason for 
believing that the true series sum exhibits the same pole at $z = 1/R_1$ that characterises 
the $[0 | 1]$ approximant.  On the other hand, if estimated values of $R_{N+1}$ from $[N-M | M]$ Pad\'e 
approximants grow progessively more accurate with increasing $N$, then properties (such as the 
whether the first positive pole precedes the first positive zero) shared by such approximants 
are likely to characterise the series itself.
Of particular interest are the ``APAP'' algorithms 
\cite{MIG}
\begin{equation}
R_3^{APAP} = 2R_2^3 / \left[ R_1^3 + R_1 R_2 \right],
\end{equation}
\begin{equation}
R_4^{APAP} = \frac{ \{R_3^2 (R_2^3 + R_1 R_2 R_3 - 2 R_1^3 R_3) \}}
{\{R_2 (2 R_2^3 - R_1^3 R_3 - R_1^2 R_2^2)\}},
\end{equation}
obtained from assuming that $[N | 1]$ approximants grow progressively accurate in predicting $R_{N+2}$ 
as $N$ increases: 
\footnote{Eq. (2.9) is the M = 1 case of the ref. \cite{IJJ} formula $\frac{
R_{N+M+1}^{[N|M]} - R_{N+M+1}^{true}}{R_{N+M+1}^{true}} = -\frac{M! A^M}{[N+M+aM+b]^M}$ with
$k_1 = -A$ and $k_2 = -A(a+b)$  The formula (2.7) is explicitly derived
in ref. \cite{ACE} via $[0|1]$ and $[1|1]$ approximants from (2.9), with
the subleading error coefficient $k_2$ chosen to be zero, as in
\cite{JJS}.  The formula (2.8) is derived in ref. \cite{MIG} via
$[0|1]$, $[1|1]$, and $[2|1]$ approximants, and is equivalent to the
prediction procedure described in \cite{IJJ}.}
\begin{equation}
\frac{R_{N+2}^{[N|1]} - R_{N+2}^{true}}{R_{N+2}^{true}} = -
\frac{k_1}{N+1+k_2 / k_1} = -\frac{k_1}{N+1} + \frac{k_2}{(N+1)^2} + ...
\end{equation}
In (2.9), $k_1$ and $k_2$ are leading and next-to-leading constants determined empirically from 
the error in $[N-1 | 1]$ and $[N-2 | 1]$ approximants in predicting ``known'' series coefficients 
$R_{N+1}$ and $R_N$,  respectively.  

	The formulae (2.7) and (2.8) are fully consistent with geometric series, such 
as the one characterising the perturbative NSVZ $\beta$-function (2.3).  If $R_1 = r$ and 
$R_2 = r^2$, it is evident from (2.7) that $R_3^{APAP} = r^3$. The situation is slightly more 
subtle for (2.8); if  $R_1 = r$, $R_2 = r^2$, and $R_3 = r^3$, $R_4^{APAP}$ is indeterminate.  
However, if $R_1$, $R_2$, and $R_3$ differ 
infinitesimally from geometric series values; {\it i.e.,} if
\begin{equation}
R_1 = r(1+\epsilon_1), \; \; R_2 = r^2(1+\epsilon_2), \; \; R_3 = r^3 (1+\epsilon_3),
\end{equation}
one easily sees from (2.8) that $R_4^{APAP} \rightarrow r^4$ as $\epsilon_{1,2,3} \rightarrow 0$:
\begin{eqnarray}
R_4^{APAP} & = & \frac{ \{r^6(1+\epsilon_3)^2 \left[ -5\epsilon_1 + 4\epsilon_2 - \epsilon_3 + {\cal{O}}(\epsilon_k^2)\right]\}}
{\{ r^2(1+\epsilon_2) \left[ -5\epsilon_1 + 4\epsilon_2 - \epsilon_3 + {\cal{O}}(\epsilon_k^2)\right]\}} \nonumber \\
& = & r^4(1+2\epsilon_3 - \epsilon_2 + {\cal{O}}(\epsilon_k^2)).
\end{eqnarray}
Thus the APAP algorithms (2.7) and (2.8) anticipate the geometric series evident 
from the first three or four known terms of (2.3) for the NSVZ $\beta$-function, and would 
therefore lead one to anticipate for that series a geometric-series pole at $z = 1/6$. 
This is born out by 
the $[2 | 1]$, $[1 | 2]$, and $[0 | 3]$ approximants to the ``known'' series terms 
$S = 1 + R_1 z + R_2 z^2 + R_3 z^3$ whose coefficients $\{R_1, R_2, R_3\}$ are given by 
(2.10):
\begin{equation}
S^{[2|1]} = \frac{1 + zr \left[\epsilon_1 - \epsilon_3 + {\cal{O}}(\epsilon_k^2)\right] + z^2 r^2 \left[\epsilon_2-\epsilon_1-\epsilon_3 + {\cal{O}}(\epsilon_k^2)\right]}
{1 - zr \left[ 1+\epsilon_3 - \epsilon_2 + {\cal{O}}(\epsilon_k^2)\right]},
\end{equation}
\begin{eqnarray*}
S^{[1|2]}  =  \frac{ \{1+zr\left[\epsilon_1 - 2\epsilon_2 + \epsilon_3 +
{\cal{O}}(\epsilon_k^2)\right] / \left[2 \epsilon_1 - \epsilon_2 +
{\cal{O}}(\epsilon_k^2)\right]\} }{ \left\{ 1+zr \frac{[\epsilon_3 -
\epsilon_1 - \epsilon_2 + {\cal{O}}(\epsilon_k^2)]}{[2\epsilon_1 -
\epsilon_2 + {\cal{O}}(\epsilon_k^2)]} + z^2 r^2 \frac{[2 \epsilon_2 -
\epsilon_1 - \epsilon_3 + {\cal{O}}(\epsilon_k^2)]}{[2\epsilon_1 -
\epsilon_2 + {\cal{O}}(\epsilon_k^2)]} \right\} } 
\end{eqnarray*}
\begin{equation}
 =  \frac{ \{\epsilon_1 (2+rz) - \epsilon_2 (1+2rz) + \epsilon_3 rz +
{\cal{O}}(\epsilon_k^2) \} }{\{\epsilon_1(2-rz-r^2z^2) -
\epsilon_2(1+rz-2r^2z^2) + \epsilon_3 (rz - r^2 z^2) +
{\cal{O}}(\epsilon_k^2) \} },
\end{equation}

\begin{equation}
S^{[0|3]} = \frac{1}{1-rz(1+\epsilon_1) + r^2 z^2 (2\epsilon_1 -
\epsilon_2) + r^3z^3 (-2\epsilon_1 + \epsilon_2 - \epsilon_3) +
{\cal{O}}(\epsilon_k^2)}
\end{equation}

\noindent All three expressions reduce to $1/(1-rz)$ in the limit 
$\epsilon_{1,2,3} \rightarrow 0$.  Moreover, if (2.10) is augmented to include $R_4 = r^4(1+\epsilon_4)$,
as anticipated by (2.11),
one can similarly show via formulae from Section 2 of \cite{CEM} that $S^{[3|1]}$, 
$S^{[2|2]}$, $S^{[1|3]}$ and $S^{[0|4]}$ determined from the ``known'' series terms 
$1+R_1 z + R_2 z^2 + R_3 z^3 + R_4 z^4$ all reduce to $1 / (1-rz)$ in the limit $\epsilon_{1,2,3,4} \rightarrow 0$,
thereby reproducing the single pole behaviour of an underlying geometric series,
such as the $r=6$ geometric series characterizing the NSVZ $\beta$-function (2.1).

It should be noted that this $1 / (1-rz)$ behaviour also characterizes
the $\epsilon_k \rightarrow 0$ limit of the lower-approximant expressions 
$S^{[0|1]}$, $S^{[1|1]}$ and $S^{[0|2]}$ [as obtained from only $R_1 = r(1+\epsilon_1)$
 and $R_2 = r^2(1+\epsilon_2)$].  Thus for $N+M \leq 4$ and $M \neq 0$ [the trivial $M = 0$ case
corresponds to the truncated series itself], {\it all} $[N|M]$
approximants  constructed from the first $N+M+1$ terms of a truncated
series $\sum_{k=0}^{N+M} R_k z^k$ reduce ultimately to the geometric
series sum $1 / (1-rz)$ as $R_k \rightarrow r^k \; \; (k \leq N + M)$.\footnote{The 
restriction $N+M \leq 4$ for the degree
of the truncated series is likely unnecessary; we have
not examined any $N+M > 4$ approximants, but suspect the result stated
here to be a general one for the coefficients $R_k$ of any truncated series.}  
Note also that any other departures from a single pole at
$z = 1/r$ ({\it e.g.}, additional zeros or poles) are seen to vanish from
these $[N|M]$ approximants as $R_k \rightarrow r^k$ for $k \leq N + M$.

These conclusions are clearly applicable to the set of $[N + M]$
approximants $(N + M \leq 4)$ constructed from truncations of the NSVZ
$\beta$-function series (a geometric series with $r = 6$), and
demonstrate the validity of such approximants in predicting the pole at
$1/6$ arising from the infinite series sum, as well as in {\it not} predicting further
spurious poles or zeros.  In the section that follows, we will utilize
the same $(N + M \leq 4)$ set of Pad\'e approximants to the DRED
$\beta$-function series for supersymmetric gluodynamics in order to
identify any common features (zeros or poles) that may be indicative of
the true infrared content of the DRED $\beta$-function series sum.

\section*{3.  Pad\'e-Approximants and DRED Supersymmetric Gluodynamics} 

\renewcommand{\theequation}{3.\arabic{equation}}
\setcounter{equation}{0} 

	The DRED $\beta$-function (2.4) is not a geometric series.  The application of 
Pad\'e-approximant methods to that $\beta$-function is predicated on the idea that such approximants 
grow progressively more accurate in predicting next-order terms, as in the error formula 
(2.9). This behaviour seems to be born out by the known coefficients appearing in (2.4).  
The APAP estimate via (2.7) for $R_3$, based on $R_1 = 6$ and $R_2 = 63$, is $R_3^{APAP} = 841.9$,
\footnote{This estimate is virtually the same as that of ref. \cite{JJS}
for the $n_f = 0$, $N_c = 3$ case, following a procedure also based on
the error formula (2.9) with $k_2 = 0$.  The $\alpha = 2.4$ estimate
derived in ref. \cite{JJS} implies that $\beta_3 = 486(1+6\alpha) =
7484$ and that $R_3 \equiv \beta_3/\beta_0 = 832$.} which is in surprisingly good 
agreement with $R_3 = 918$, supersymmetric gluodynamics' 
true $R_3$ value in the DRED scheme \cite{IJD}.   By contrast, the naive estimate for $R_3$ based on 
the $[1 | 1]$ approximant reproducing $1 + 6y + 63y^2$ is $R_3^{[1 | 1]} = (63)^2 /6 = 661.5$.  
Thus, the success of (2.7) [which devolves from (2.9)] in predicting $R_3$ more accurately 
than the $[1 | 1]$ 
approximant itself may be seen as evidence for the progressive increase in accuracy of 
$[N-M | M]$ approximants in reproducing the DRED $\beta$-function as $N$ increases.

	The known terms (2.4) are sufficient to determine $[2 | 1]$- and $[1 | 2]$-approximant 
versions of the DRED $\beta$-function:
\begin{equation}
\beta^{[2|1]}(y) = -9y^2 \left[ \frac{1 - 8.5714y - 24.4286y^2}{1 -
14.5714y} \right]
\end{equation}
\begin{equation}
\beta^{[1|2]}(y) = -9y^2 \left[ \frac{1 - 14y}{1 - 20y + 57y^2} \right].
\end{equation}
Both of these approximations to the true DRED $\beta$-function necessarily have an ultraviolet 
fixed point at $y = 0$.  However, their infrared behaviour, if extractable at all, should 
manifest itself in a positive zero or pole common to the asymptotically free phase of both 
expressions. We find that both approximants are consistent with infrared dynamics governed 
by an infrared-attractor pole, as is the case for the NSVZ $\beta$-function.  For 
$\beta^{[2 | 1]}$, a pole at $y = 0.06863$ is seen to precede the first positive 
numerator zero (0.09236). Similarly, the 
first positive pole of $\beta^{[1 | 2]}$ is seen to occur at $y = 0.0604$, which precedes 
the positive 
numerator zero at $y = 0.07143$. Thus in both cases, the pole controls the infrared dynamics 
of the asymptotically-free phase in much the same way as the NSVZ $\beta$-function pole. \footnote{The 
larger numerator zero, if meaningful at all, corresponds to an ultraviolet fixed point of a 
subsequent (non-asymptotically free) phase of the couplant.   See the discussion of subsequent
positive zeros/poles for toy models discussed in section 2 of ref. \cite{CEM}.} 
Such a positive pole also characterizes the $[0|3]$-approximant to the DRED $\beta$-function,
as determined from the full set of known terms in the series (2.4), as well as $[1|1]$ and $[0|2]$ approximants 
constructed from the first three terms of that series. 

	One can consider higher Pad\'e-approximants by using (2.8) to obtain an estimate 
of $R_4$.  Using $R_1 = 6$ , $R_2 = 63$, and $R_3 = 918$, one sees from (2.8) that $R_4^{APAP} = 16,874$, 
corresponding to $\beta_4^{DRED} = 151,870$.\footnote{This estimate is
larger than the estimate $\beta_4 = 113,000$ of \cite{IJJ},
because of that latter estimate's explicit use of $\alpha=2.4$ as an input parameter.  If
$\alpha$ is taken to be 2.4 instead of its true value of 8/3, then $R_3
= 832$ (see Footnote 4).  Eq. (2.8) would then imply that $R_4^{APAP} =
12,700$ and that $\beta_4 = 9R_4^{APAP} = 114,000$, consistent with the
estimate in \cite{IJJ}.}
Given this estimate for $R_4$, it is straightforward to 
generate $[3 | 1]$, $[2 | 2]$, and $[1 | 3]$ Pad\'e-approximant versions of the DRED  
$\beta$-function (formulae for constructing these approximants from series are explicitly 
given in \cite{CEM}):
\begin{equation}
\beta^{[3|1]}(y) = -9y^2 \left[ \frac{1-12.38y-47.29y^2-240.0y^3}{1-
18.38y} \right]
\end{equation}
\begin{equation}
\beta^{[2|2]}(y) = -9y^2 \left[ \frac{1-22.21y+36.93y^2}{1-
28.21y+143.2y^2} \right]
\end{equation}
\begin{equation}
\beta^{[1|3]}(y) = -9y^2 \left[ \frac{1-19.57y}{1-
25.57y+90.41y^2+150.4y^3} \right].
\end{equation}
As before, the first positive pole of each of these functions precedes the first positive 
zero (Table 1), consistent with the pole-dominated infrared dynamics of the NSVZ $\beta$-function's 
asymptotically-free phase. 

	Table 1 tabulates the first positive pole and zero of {\it every} Pad\'e-approximant version 
of the DRED $\beta$-function obtainable from known $(R_1 = 6 , R_2 = 63, R_3 = 918)$ and APAP-estimated 
$(R_4^{APAP} = 16,874)$ values of the leading series coefficients. Every approximant considered 
exhibits a positive pole which precedes any positive zeros of the approximant. Such an 
ordering implies that the first positive pole is necessarily associated with the infrared
limit of the asymptotically-free phase of the couplant.  
Indeed, the leading positive Pad\'e-approximant zero can be associated with an infrared 
fixed-point for the asymptotically-free phase only if it {\it precedes} any poles of the same 
approximant \cite{MIG}. This is clearly {\it not} the case for DRED $\beta$-function, which appears (from all of 
its approximant versions) to exhibit the same infrared-attractor pole dynamics as the 
NSVZ $\beta$-function for supersymmetric gluodynamics.

\section*{4. Discussion}

\renewcommand{\theequation}{4.\arabic{equation}}
\setcounter{equation}{0}

	For the asymptotically-free phase of supersymmetric gluodynamics, the existence of 
an infrared-attractor pole within every Pad\'e-approximant version of the DRED $\beta$-function is 
strong evidence that such a pole is not an artifact.  Thus, it appears that DRED and NSVZ lead 
to infrared dynamics that are qualitatively equivalent. However, it must also be noted from 
Table 1 that the DRED version of this infrared-attractor pole appears to be {\it smaller} than 1/6, 
the value of the NSVZ pole. [The value of 1/6 in Table 1 obtained from the $[0 | 1]$ approximant to the DRED  
$\beta$-function is irrelevant, as the exact same approximant characterizes the true NSVZ $\beta$-function, as 
well as the $[0|1]$ approximant generated by its leading series terms.] There is, of course, no reason 
why the poles for the two schemes {\it should} have the same value; it is evident from (1.8) that 
corresponding values of the couplant in the two schemes are inequivalent. 

	Nevertheless, if NSVZ and DRED values of the couplant are equal at some (large) finite 
value of $\mu$, one would anticipate from the $\beta$-function series (2.3) and (2.4)
that any particular NSVZ couplant value at a lower value of $\mu$ scheme should correspond 
to a {\it larger} couplant value in the DRED scheme, since the $\beta$-function coefficients of the latter 
are term-by-term larger than the $\beta$-function coefficients of the former. This anticipated 
inequality is reflected in the relation (1.8), which clearly generates DRED couplants (y) 
that are larger than their corresponding NSVZ couplants $(x)$. Although the value $x = 1/6$ is 
very likely outside the radius of convergence of the series in (1.8), the above observations 
seem indicative of a fundamental inconsistency in the infrared content of  DRED and NSVZ 
supersymmetric gluodynamics. Specifically, if the infrared-attractor poles of the two schemes 
are in correspondence via a redefinition of the couplant from one scheme to another, then 
(1.8) suggests that the DRED pole should be {\it larger} than the NSVZ pole.

	In fact, it may indeed be possible for a given benchmark value of the NSVZ couplant to correspond to a 
{\it smaller} value of the DRED couplant within supersymmetric gluodynamics.  
For the perturbative regime, one can demonstrate that (1.8) is not necessarily the most general 
possible relationship between couplants in the asymptotically-free phases of the two schemes,
as defined by their respective $\beta$-functions.  Consider first a general 
parametrisation of the perturbative relation between the two couplants that is consistent 
with both couplants being in their asymptotically-free phases [{\it i.e.,} when $x \rightarrow 0$, 
one must also require that $y \rightarrow 0$]:
\begin{equation}
y = x(1+\alpha x + \beta x^2 + \gamma x^3 + ...)
\end{equation}
One sees from the series expansion (2.3) of the NSVZ $\beta$-function that 
\begin{eqnarray}
\beta^{DRED} (y) & = & \mu^2 \frac{dy}{d\mu^2} = \mu^2 \frac{dx}{d
\mu^2} \left[ 1 + 2 \alpha x + 3 \beta x^2 + 4\gamma x^3 + ... \right]
\nonumber \\
& = & -9x^2 \left[ 1 + (6+2 \alpha) x + (36+3 \beta + 12\alpha) x^2
\right. \nonumber \\
& + & \left. (216 + 4\gamma + 72\alpha + 18\beta) x^3 + ... \right]
\end{eqnarray}
However, if one substitutes (4.1) into (2.4), the series expansion for the DRED  
$\beta$-function, one finds that 
\begin{eqnarray}
\beta^{DRED} (y) & = & -9x^2 \left[ 1+(6+2\alpha) x + (63+18\alpha +
\alpha^2 + 2\beta) x^2 \right. \nonumber \\
& + & \left. (918 + 252\alpha + 18\alpha^2 + 2\alpha \beta + 18 \beta + 2\gamma)
x^3 + ... \right]
\end{eqnarray}
Comparing the final line of (4.2) to  (4.3), one finds that 
\begin{equation}
\beta=27+6\alpha + \alpha^2, \; \; \gamma = 351 + 117\alpha + 15\alpha^2
+ \alpha^3,
\end{equation}
and that
\begin{equation}
y = x \left[ 1+\alpha x + (27+6\alpha+\alpha^2) x^2 + (351+117\alpha+15\alpha^2 + \alpha^3) x^3 + ... \right].
\end{equation}
In the absence of any further input information,
the value of the coefficient $\alpha$  in(4.1) remains {\it undetermined}, a consequence of the fact that DRED and 
NSVZ $\beta$-functions agree to two-loop order.  If we choose $\alpha = 0$, we recover the relation 
(1.8), which implies $y > x$ for corresponding perturbative couplant values in DRED and NSVZ, 
respectively.  The choice $\alpha = 0$ is intuitive, in that it forces corresponding NSVZ and DRED 
couplants to agree to two orders, as is the case for the $\beta$-functions themselves of both schemes. However, 
it appears that the only methodological constraint one is forced to impose is that both couplants be in their 
asymptotically free phases, as remarked above, and the form of (4.1) is consistent 
with this requirement for {\it any} finite value of $\alpha$.   We see that if $\alpha$ is sufficiently negative, 
the terms in the series (4.5) may alternate in sign.  For example, 
if we choose $\alpha = -6$, we see from the four terms of the series (4.5) that the NSVZ pole 
at $x = 1/6$ corresponds to a DRED value $y = 5/48$. If (for this value of $\alpha$) the series 
continues term-by-term to decrease in magnitude and alternate in sign, 
the three- and four-term partial-sum values $1/8$ and $5/48$ become upper and lower bounds 
on the DRED value corresponding to the NSVZ pole at $x = 1/6$. 

	The above scenario is, of course, a contrived one.  The point we wish to make, however, 
is that there appears to be no real restriction on the relative sizes of corresponding DRED and NSVZ 
values of the couplant within supersymmetric gluodynamics in the perturbative region. 
Thus, we conclude that 

\begin{quote}1)  the Pad\'e-approximant analysis indicates the existence of an 
infrared-attractor pole within DRED supersymmetric gluodynamics that is smaller than the 
infrared-attractor pole already known to characterise NSVZ supersymmetric gluodynamics 
\cite{KS}, and that 
\end{quote}

\begin{quote} 2)  the relative sizes of these poles are {\it not necessarily} incompatible with the 
order-by-order correspondence between couplants of the two schemes.
\end{quote}

\section*{Acknowledgments}

I am indebted to V. A. Miransky for suggesting a Pad\'e-approximant discussion of NSVZ 
supersymmetric gluodynamics, and to F. A. Chishtie and T. G. Steele
for numerous discussions related to the material presented here. I am also grateful for 
two weeks of fruitful interaction 
on Pad\'e-approximant methods with Mark Samuel immediately prior 
to his untimely passing. Research support from the Natural Sciences and Engineering Research 
Council of Canada is also gratefully acknowledged.

\begin{table}
\begin{center}
\begin{tabular}{cccc}
Approximant & Inputs & First Positive Pole & First Positive Zero\\
\hline\\
$[0|1]$ & $R_1$ & $1/6$ & ---\\
$[1|1]$ & $R_{1,2}$ & $2/21$ & $2/9$\\
$[0|2]$ & $R_{1,2}$ & $1/9$ & ---\\
$[2|1]$ & $R_{1,2,3}$ & 0.0686 & 0.0924\\
$[1|2]$ & $R_{1,2,3}$ & 0.0604 & 0.0714\\
$[0|3]$ & $R_{1,2,3}$ & 0.0773 & ---\\
$[3|1]$ & $R_{1,2,3,4}$ & 0.0544 & 0.0617\\
$[2|2]$ & $R_{1,2,3,4}$ & 0.0464 & 0.0490\\
$[1|3]$ & $R_{1,2,3,4}$ & 0.0479 & 0.0511\\
$[0|4]$ & $R_{1,2,3,4}$ & 0.0745 & ---\\
\end{tabular}
\caption{The first positive pole and first positive zero of all possible Pad\'e approximant versions
of the DRED $\beta$-function are tabulated for $R_1 = 6$, $R_2 = 63$, $R_3 = 918$,
and $R_4 = 16,874$, as discussed in the text.  Series parameters
used to determine each approximant are listed in the second column.
Every approximant exhibits a positive pole which precedes any positive
zeros of the same approximant.}
\end{center}
\end{table}


\begin{thebibliography}{99}

\bibitem{NSV}Novikov V, Shifman M, Vainshtein A and Zakharov V,
1983 {\it Nucl. Phys.} B {\bf 229} 381
\bibitem{ACV}Avdeev L N, Chochia G A and Vladimirov A A 1981 {\it Phys. Lett.} B {\bf 105} 272
\bibitem{DRT}Jones D R T 1983 {\it Phys. Lett.} B {\bf 123} 45
\bibitem{DJL}Jones D R T and Leveille J 1982 {\it Nucl. Phys.} B {\bf 206} 473
\bibitem{KS}Kogan I I and Shifman M 1995 {\it Phys. Rev. Lett.} {\bf 75} 2085
\bibitem{JJN}Jack I, Jones D R T and North C G 1997 {\it Nucl. Phys.} B {\bf 486} 479
\bibitem{IJD}Jack I, Jones D R T and Pickering A  1998 {\it Phys. Lett.} B {\bf 435} 61
\bibitem{BG}Baker G and Graves-Morris P 1981 {\it Pad\'e-Approximants}
[Vol. 13 of {\it Encyclopedia of Mathematics and its Applications}] (Reading,
MA: Addison-Wesley) pp 48-57
\bibitem{CEM}Chishtie F A, Elias V, Miransky VA  and Steele T G 2000
{\it Prog. Theor. Phys.} {\bf 104} (to appear: Los Alamos Archive Preprint hep-ph/9905291).
\bibitem{MIG}Elias V, Steele T G, Chishtie F, Migneron R and Sprague K  1998 {\it Phys. Rev.}
 D {\bf 58}  116007
\bibitem{IJJ}Ellis J, Jack I, Jones D R T, Karliner M and Samuel M A  1998 {\it Phys. Rev.}
D {\bf 57} 2665
\bibitem{ACE}Ahmady M R, Chishtie F A, Elias V and Steele T G 2000 {\it Phys. Lett}. B {\bf 479} 201
\bibitem{JJS}Jack I, Jones D R T and Samuel M A 1997 {\it Phys. Lett.} B {\bf 407} 143
\end{thebibliography}
\end{document}